\documentstyle[12pt,epsfig]{article}
\begin{document}

\begin{center}

\vspace{3cm}

{\large\bf{ Longitudinal top quark polarization }}

\vspace{2cm}

{\bf Ma{\l}gorzata Awramik$^{a}$, Marek Je\. zabek$^{b}$}

\vspace{0.4cm}
{\it  
	$^a$Department of Field Theory and Particle Physics,\\ 
        University of  Silesia, \\
        Uniwersytecka 4, PL-40007 Katowice, Poland \\ 
   \vspace{0.4cm}
	$^b$Institute of Nuclear Physics, \\
        Kawiory 26 a, PL-30055 Cracow, Poland\\
        }
\end{center}

\vspace{1cm}

\begin{abstract}
Longitudinal polarization of the top quark, averaged over the production
angle, is discussed for the top quark produced in $e^+e^-$
annihilation near its production threshold.
It is demonstrated that Coulomb type corrections and rescattering corrections
are important. They change considerably measurable quantities and should be taken into 
account in phenomenological analysis.  

\vspace{0.5cm}
\hspace{-0.7cm}
PACS numbers: 13.88.+e, 14.65.Ha

\end{abstract}

\newpage
\section{ Introduction}  

The heaviest of all known elementary particles,
the top quark, is likely to give us
exciting insight into the electroweak symmetry breaking sector of the
Standard Model (SM), into QCD dynamics at short distances and maybe
even into physics beyond the SM.
Precise measurements not only of the total top quark production cross
section, but also of the top quark momentum distributions and polarizations are
planned to be carried out at a future linear electron-positron collider.
Of particular interest is the subject of top-antitop production close
to threshold. For the large width of the top quark $\Gamma_t$ ~\cite{bdkkz} 
a formalism suitable for this region, where the strong
interaction between $t$ and $\bar t$ due to Coulomb-like gluons 
is of paramount importance,
was suggested by V.S.~Fadin and V.A.~Khoze \cite{fadin_khoze}. It is
based on solving Schr\"{o}dinger equation for Green functions 
in a non-relativistic approximation for a Coulombic potential.
The methods needed for the calculation of the total cross section were
first given in \cite{bdkkz} and \cite{strassler}. 
The differential top quark distributions were calculated independently 
in position space \cite{mursum, sphd} and by solving
Lippmann-Schwinger equations in momentum space \cite{jkt, jt93}.
The polarization of the top quark has been considered in \cite{hjkt,HJKP}.
In \cite{mpys,mjys} rescattering corrections (between $t\bar t$ and
the decay $b$ and $\bar b$ quarks) to order ${\cal O}(\alpha_s)$ were
included\footnote{For a review and references see e.g. \cite{mj95,thom, accomando}}. 

One of the interesting quantities is the top quark longitudinal polarization.
It has been claimed in \cite{vol} that it may easily
provide information on the mass and width of the top quark,
independently of Coulomb type corrections and of the running of the strong
coupling $\alpha_s$. However, in this article we demonstrate that
the helicity of the top quark significantly depends on Coulomb
type corrections. Moreover, we show that some easily measurable 
quantities related to the top net helicity strongly depend on 
rescattering corrections and therefore it requires a more careful analysis to extract 
from them some information on top quark parameters.
Although the NNLO QCD 
corrections to the total and differential cross section (excluding rescattering corrections) 
for the top quark production in 
the electron-positron annihilation are already known (see \cite{teubner_hoang} and \cite{nnlo}; 
recent results are given in  \cite{hmst}), polarization has not yet been introduced. A major problem 
is caused by the rescattering corrections since the inclusion of the interactions between relativistic 
(b quark) and non-relativistic (t quark) states is difficult within this formalism.  

In section 2 we  investigate the top quark net helicity when
rescattering corrections are neglected. 
We show that large Coulomb type corrections 
affect this observable, in contradiction to the
statements made in \cite{vol}.
For unstable quarks, the polarization is a function of the momentum and  energy 
independently\footnote{We take the energy with reference to the nominal threshold
 $E=\sqrt{s}-2m_t$}. 
To compare our results to \cite{vol} we need some averaging procedure. 
We show, however, that differences persist in two intuitively acceptable approaches.
Inclusion of the phenomenological potential $V_{JKT}$ (see \cite{jkt}) with the two loop static potential, 
running of the strong coupling constant and a Richardson-like potential for small momenta 
leads to further deviations from \cite{vol}.

In section 3, we note that the top quark average helicity is not directly observable. 
Following \cite{HJKP} we derive an analogue from the leptonic distributions, 
taking a more suitable choice of basis vectors. 
We show, however, that the latter quantities are significantly 
affected by the final state interactions. 

Our conclusions are presented in section 4. 

The Appendix contains the formulas necessary to express the observables of \cite{HJKP} 
in a basis aligned with the top quark momentum.

\section{Coulomb corrections and finite width}

We start from the  definition of the angular average of the top quark polarization 
\begin{equation}
\begin{Large}
\langle {\cal P} \rangle=\frac{\int{\cal P}\, \frac{d\sigma}{dp\, d\Omega_p}\, d\Omega_p}
                       {\int\frac{d\sigma}{dp\, d\Omega_p}\, d\Omega_p}\,
\end{Large}
\label{pola}
\end{equation}
where $\frac{d\sigma}{dp\, d\Omega_p}$ is the momentum distribution of
the top quark and ${\cal P}$ its polarization. 
The angular average of the longitudinal polarization is
\begin{equation}
\langle{\cal P}_{L}\rangle = \frac{4}{3}\, C_{\bot}\, \varphi_R
\label{sec}
\end{equation}
where \(C_{\bot}\) is a well known function of the electroweak couplings of 
the t quark to the photon and $Z^0$. In Appendix A we give a derivation of Eq.~(\ref{sec})
using the results of \cite{hjkt} and \cite{HJKP}. An explicit formula for  \(C_{\bot}\)
is given therein, see Eq.~(\ref{coefs}). The function \( \varphi_R = Re ( \varphi ) \)
originates from the interference of the $S$ and $P$ wave production
\begin{equation}
 \varphi (p,E) = \frac{ (1-4 \alpha_s /3\pi)}
             { (1-8 \alpha_s /3\pi)} 
         \frac{p}{m_t} \frac{ F^{*}(p,E)}{ G^{*}(p,E)} .
\label{phi}
\end{equation}
The functions $G$ and $F$  correspond to the respective waves
and are found numerically by solving the Lippmann-Schwinger equations.

\begin{figure}[h!]
\begin{center}
\epsfig{file=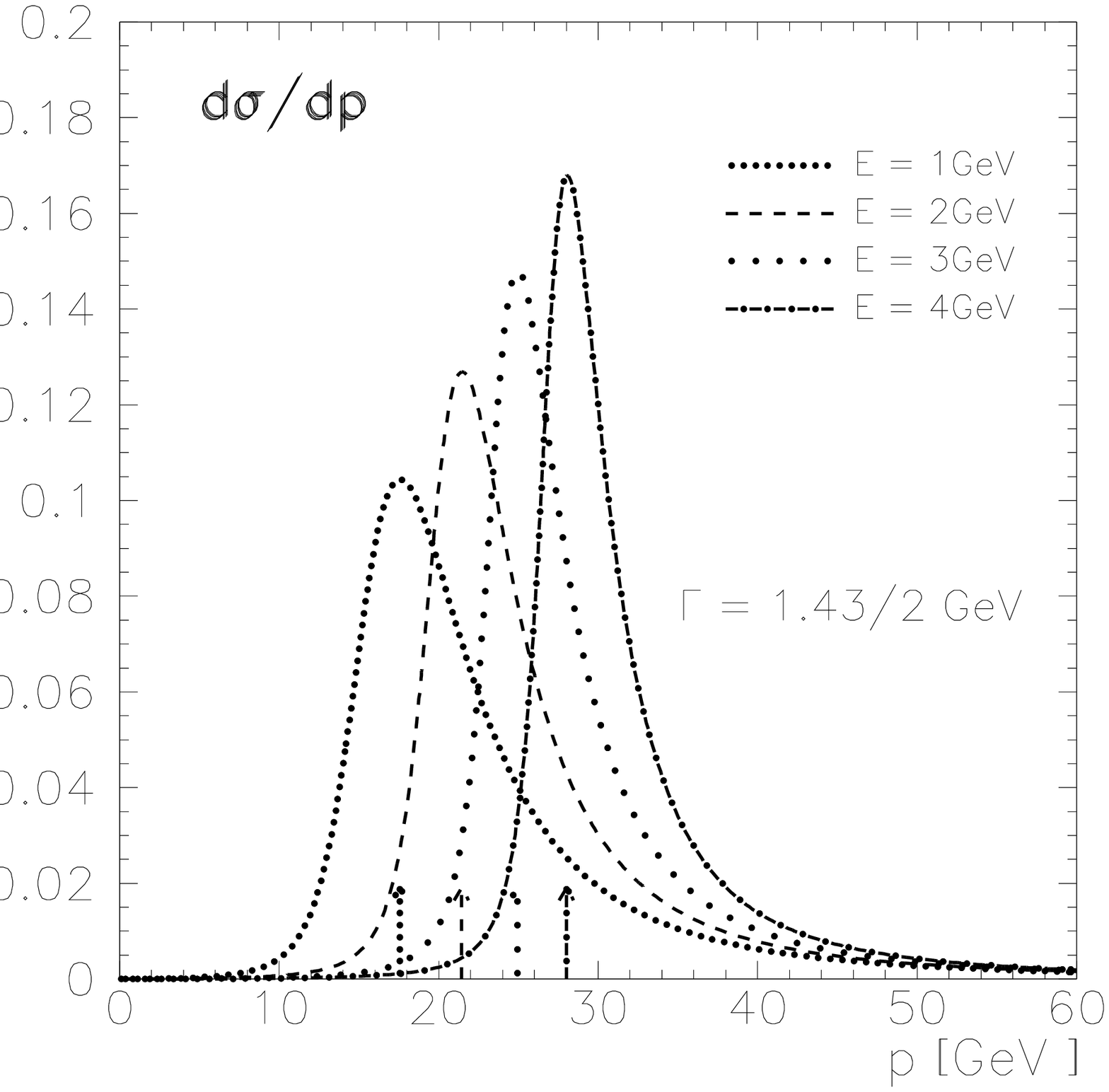,width=6.5cm,height=5.5cm}
\epsfig{file=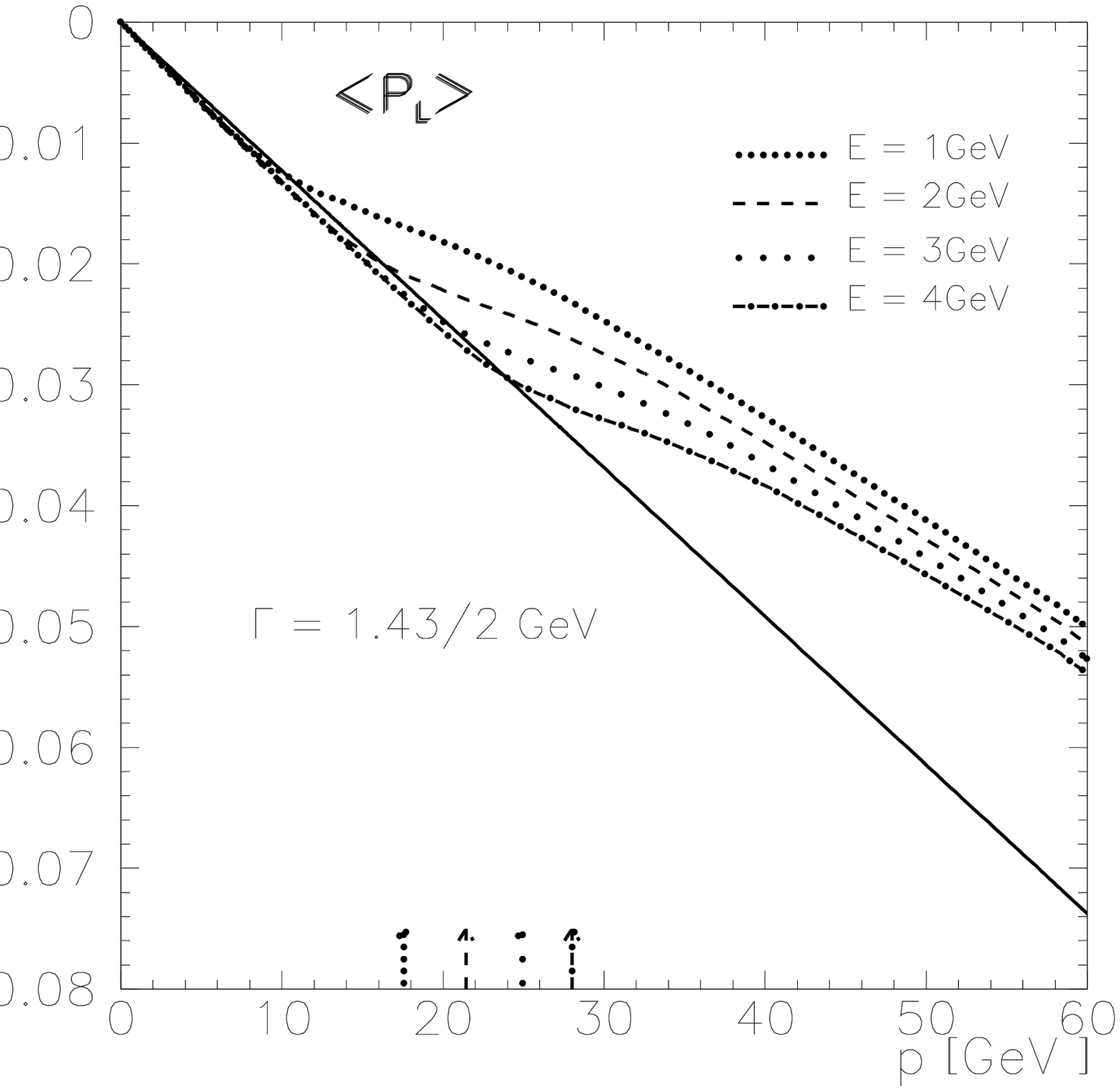,width=6.5cm,height=5.5cm}
\epsfig{file=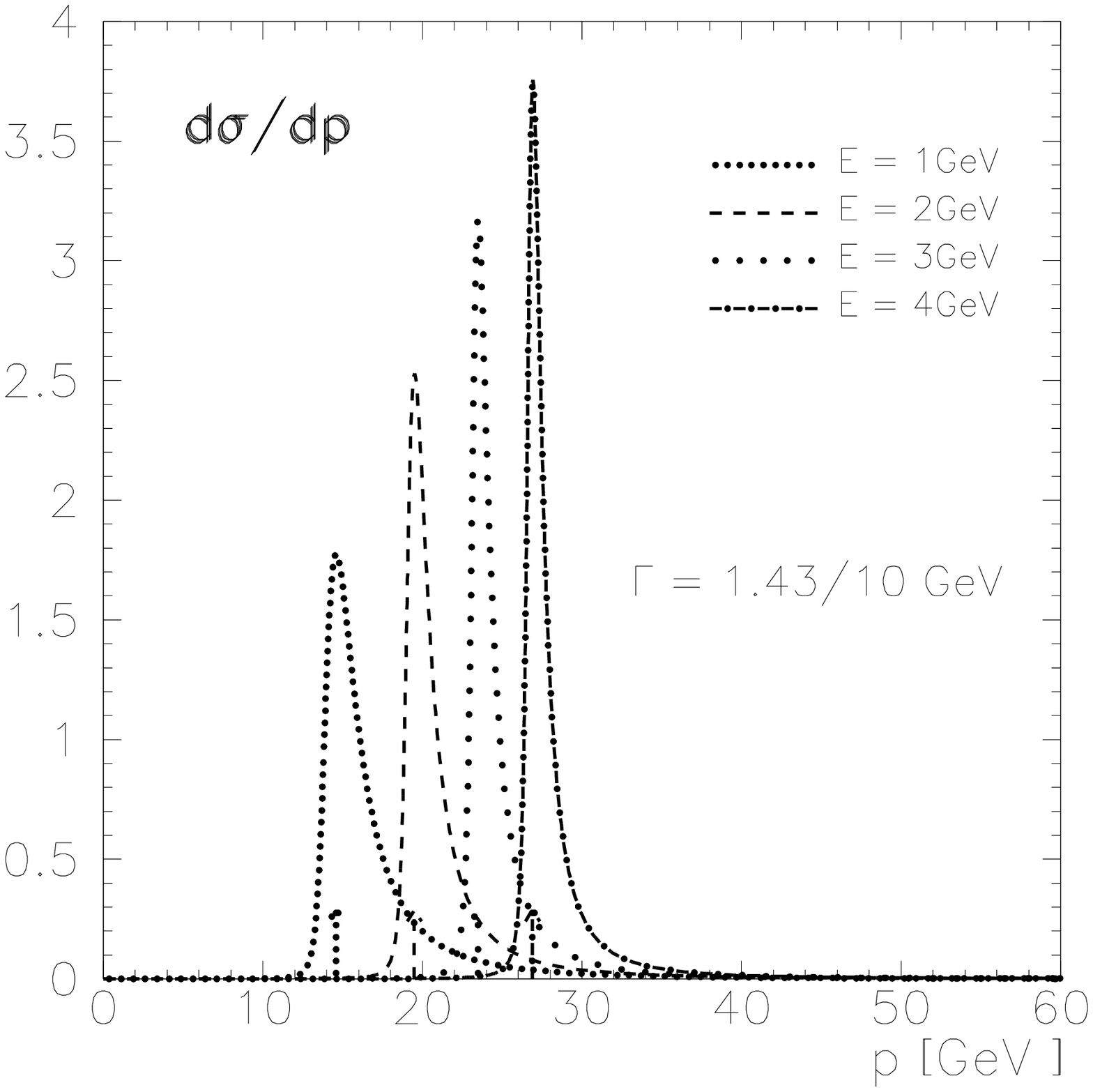,width=6.5cm,height=5.5cm}
\epsfig{file=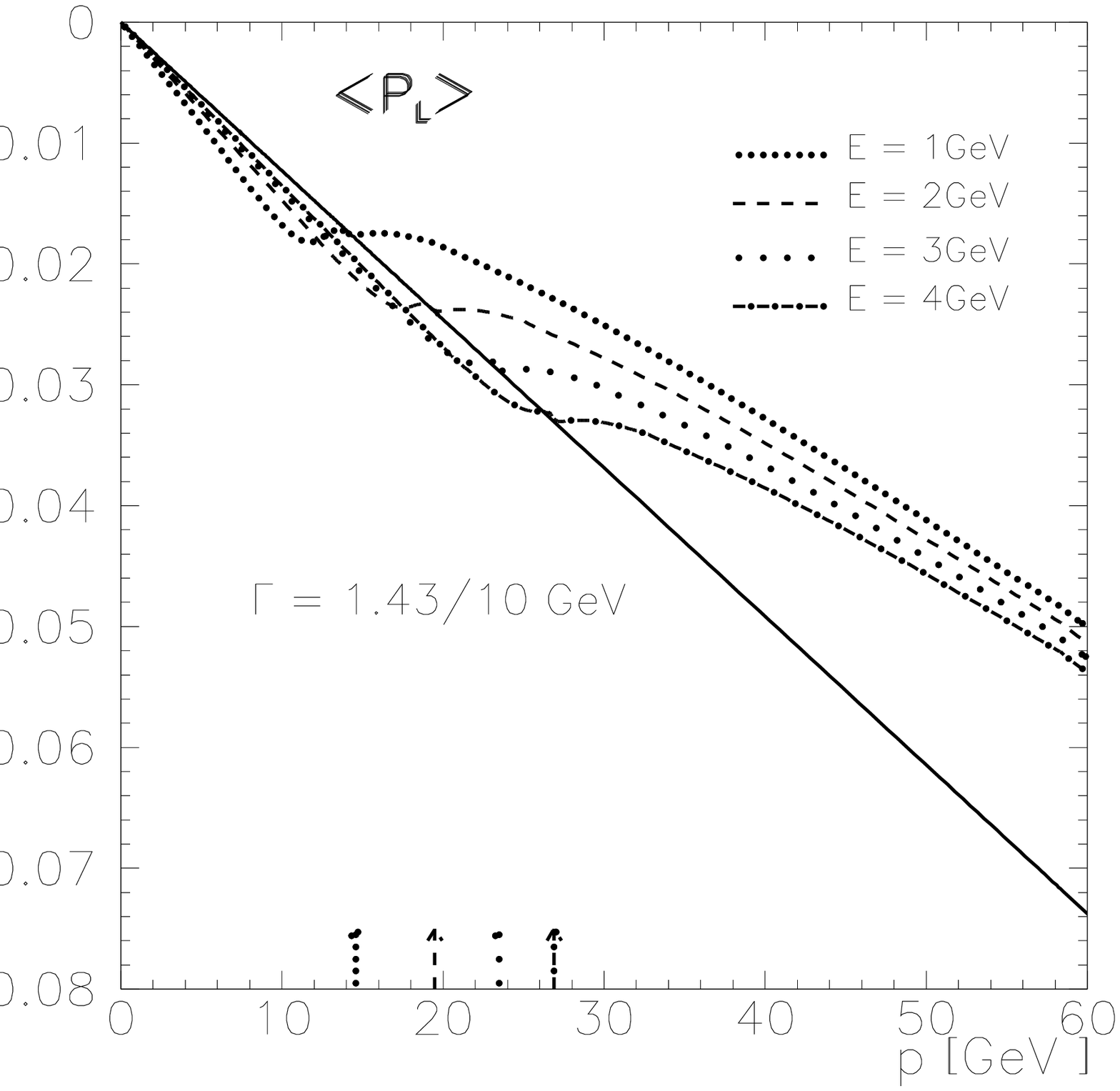,width=6.5cm,height=5.5cm}
\caption{\footnotesize \it
	On the left: top quark momentum distributions 
        $d\sigma/dp$. 
	On the right: the polarization $\langle
        {\cal P}_L\rangle$ from the $S$-$P$ wave interference term 
        shown for several different energies, 
        compared with  $\langle {\cal P}_0\rangle$ (solid lines).
        The upper plots show the dependence for an artificially small
        width $\Gamma_t \rightarrow \Gamma_t/2$, the lower are for
        $\Gamma_t \rightarrow \Gamma_t/10$. The arrows indicate the
        peak of the momentum distribution. We have chosen
        $m_t=175$~GeV, $\Gamma_t=1.43$~GeV. We use the pure Coulomb potential 
	with $\alpha_s$ fixed at average value of the peak momentum 
	$\alpha_s(p=20)=0.15$. }
\label{fig1}
\end{center}
\end{figure}

For stable and non-relativistic quarks and assuming a pure Coulomb potential $G$ and $F$
can be found analytically ~\cite{fkk}. 
Following the remarks in~\cite{fkk} and ~\cite{HJKP} we rewrite Eq.~(\ref{phi}) as
\begin{equation}
\lim_{\Gamma_t \rightarrow 0,\, E \rightarrow \frac{p^2}{m_t} }  
        \varphi_R \equiv \varphi_0 
                = \frac{ (1-4 \alpha_s /3\pi)}
                       { (1-8 \alpha_s /3\pi)} 
                   \beta\,.
\label{phi_lim}
\end{equation}
Here $ \beta=\sqrt{1-4m_t^2/s}$ is the on-shell top quark
velocity. In this limit the longitudinal top polarization is given by
\begin{equation}
\lim_{\Gamma_t \rightarrow 0,\,  E \rightarrow \frac{p^2}{m_t}}
        \langle{\cal P}_{L}\rangle
\equiv
        \langle{\cal P}_{0}\rangle
                = \frac{4}{3}\, C_{\bot}\, Re (\varphi_0)\,.
\nonumber
\label{pl_lim}
\end{equation}

In Fig.~\ref{fig1} we show how Eq.~(\ref{pola}) approximates to Eq.~(\ref{pl_lim}) as
$\Gamma_t$ decreases. Although the momentum distributions are getting narrower to simulate the Dirac delta function,
the limiting value of the polarization of Eq.~(\ref{pola}) is reproduced only in the near vicinity of the peak.

In \cite{vol}, the Eq.~(\ref{pl_lim}) was used as a starting point.
The finite width was taken into account by changing the dispersion relation
\begin{equation}
p = \sqrt{m_t \, (E + i\Gamma_t)}\,,
\label{mom}
\end{equation}
which leads to the longitudinal top polarization
\begin{equation}
\left(\langle {\cal P}_{L} \rangle \right)_S = \frac{4}{3} \, (1\,+\frac{4 \,\alpha_s\,}{\,3\,\pi\,})\,
                          C_\perp \,Re(\,\sqrt{\frac{ E+i\Gamma_t}{m_t}} \, )\,.
\label{pola4}
\end{equation}
\begin{figure}[h!]
\begin{center}
\epsfig{file=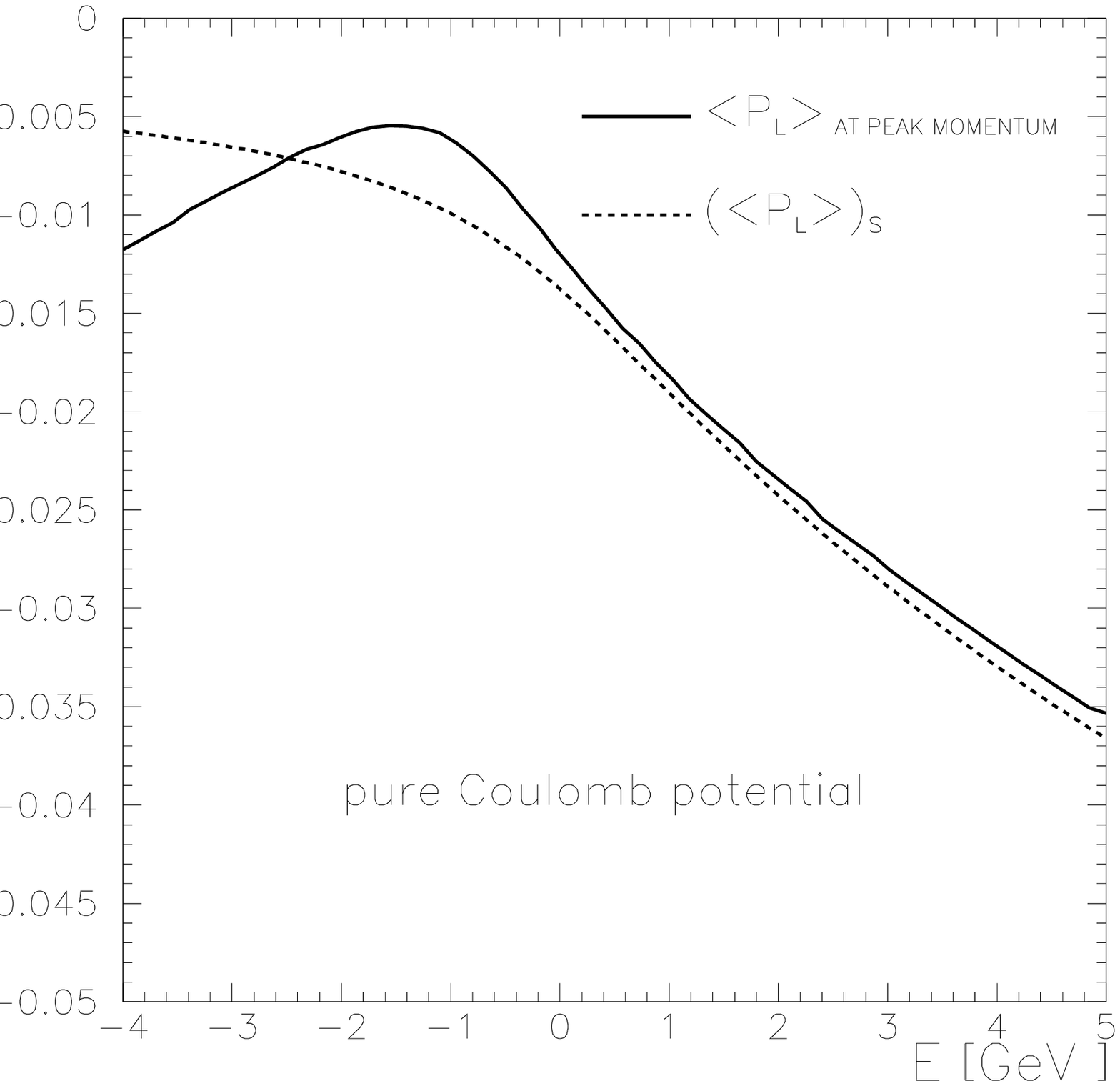,width=6.5 cm,height=6.5 cm}
\epsfig{file=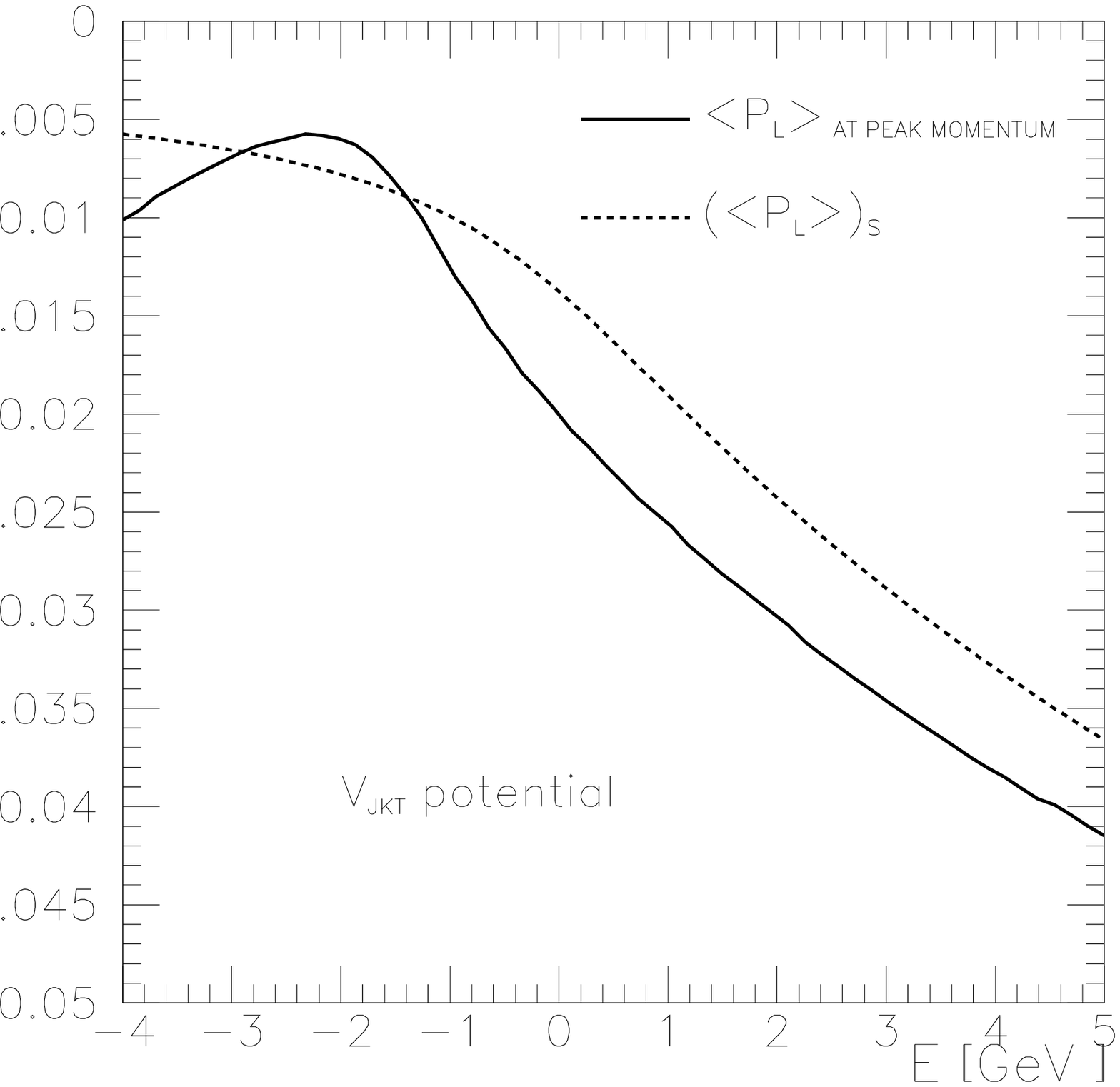,width=6.5 cm,height=6.5 cm}
\caption{\footnotesize \it  Longitudinal polarization $\langle {\cal P}_L\rangle$ (solid line) 
	taken at the peak 
        of the momentum distribution for the pure Coulomb potential (left)
        and phenomenological potential $V_{JKT}$  (right) compared with 
        $\left(\langle {\cal P}_L\rangle\right)_S$ (dashed line ).}
\label{fig2}
\end{center}
\end{figure}
In a more complete analysis the width is accounted for by momentum distributions. Therefore
a comparison with Eq.~(\ref{pola4}) requires some averaging. We shall consider two possibilities.
In Fig.~\ref{fig2} we take the value of the polarization at the peak of the momentum
distribution as function of the energy. In Fig.~\ref{fig3}, we show the average
\begin{equation}
\begin{Large}
\langle\langle {\cal P}\rangle\rangle = \frac
                {\int_{0}^{p_{max}}\,{\cal P} 
                                \, \frac{d\sigma}{dp \, d\Omega_{p}}\, d\Omega{_p}\, dp}
                {\int_{0}^{p_{max}}        
                                \, \frac{d\sigma}{dp \, d\Omega_{p}} \, d\Omega{_p}\, dp}
\end{Large}
\label{pola2}
\end{equation}
for the longitudinal part of the top polarization

\begin{equation}
\begin{Large}
\langle\langle {\cal P}_{L}\rangle\rangle = \frac
                {\int_{0}^{p_{max}} 
                         \, \langle {\cal P}_L(p,E)\rangle\, |p \, G(p,E)|^2 \, dp}
                {\int_{0}^{p_{max}}        
                         \, |p \, G(p,E)|^2 \, dp}.
\end{Large}
\label{pola3}
\end{equation}
Due to the  non-relativistic approximation employed in our calculation of the Green
functions $F$ and $G$, $\langle \langle P_L \rangle \rangle$ increases logarithmically with $p_{max}$. 
We show the results for two values of the cut off,
$p_{max}=\frac{2}{3} m_t $ and $p_{max}=\frac{1}{4} m_t $. The former cut off 
corresponds to more then 97\% of the total cross section in the energy
range between 1S peak and $E = 5 $~GeV. In the latter case about 80\%
of the total cross section is included. So our calculation should be
compared with the data sample for a corresponding cut off. 

\begin{figure}[h!]
\begin{center}
\epsfig{file=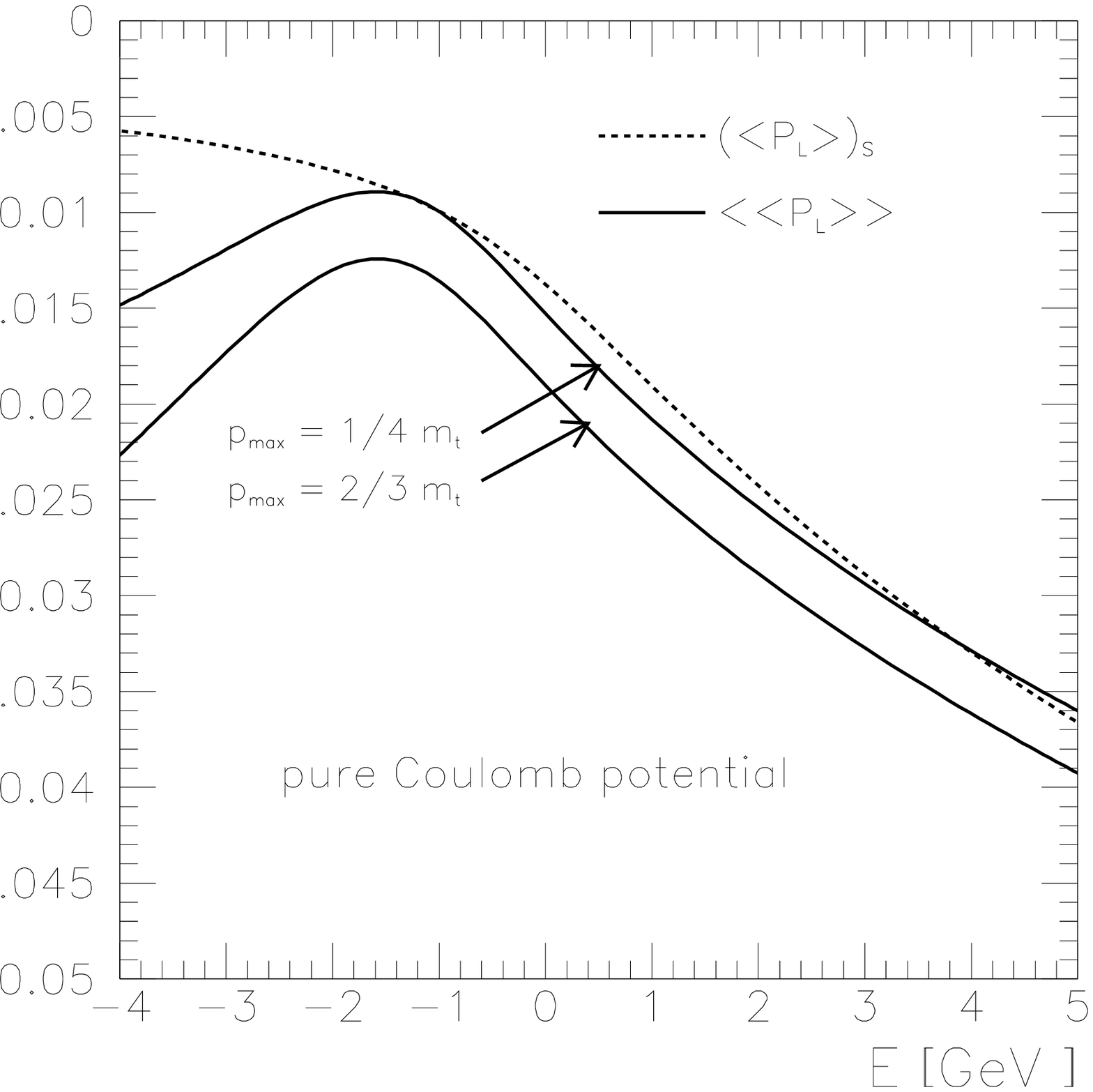,width=6.5cm,height=6.5cm}
\epsfig{file=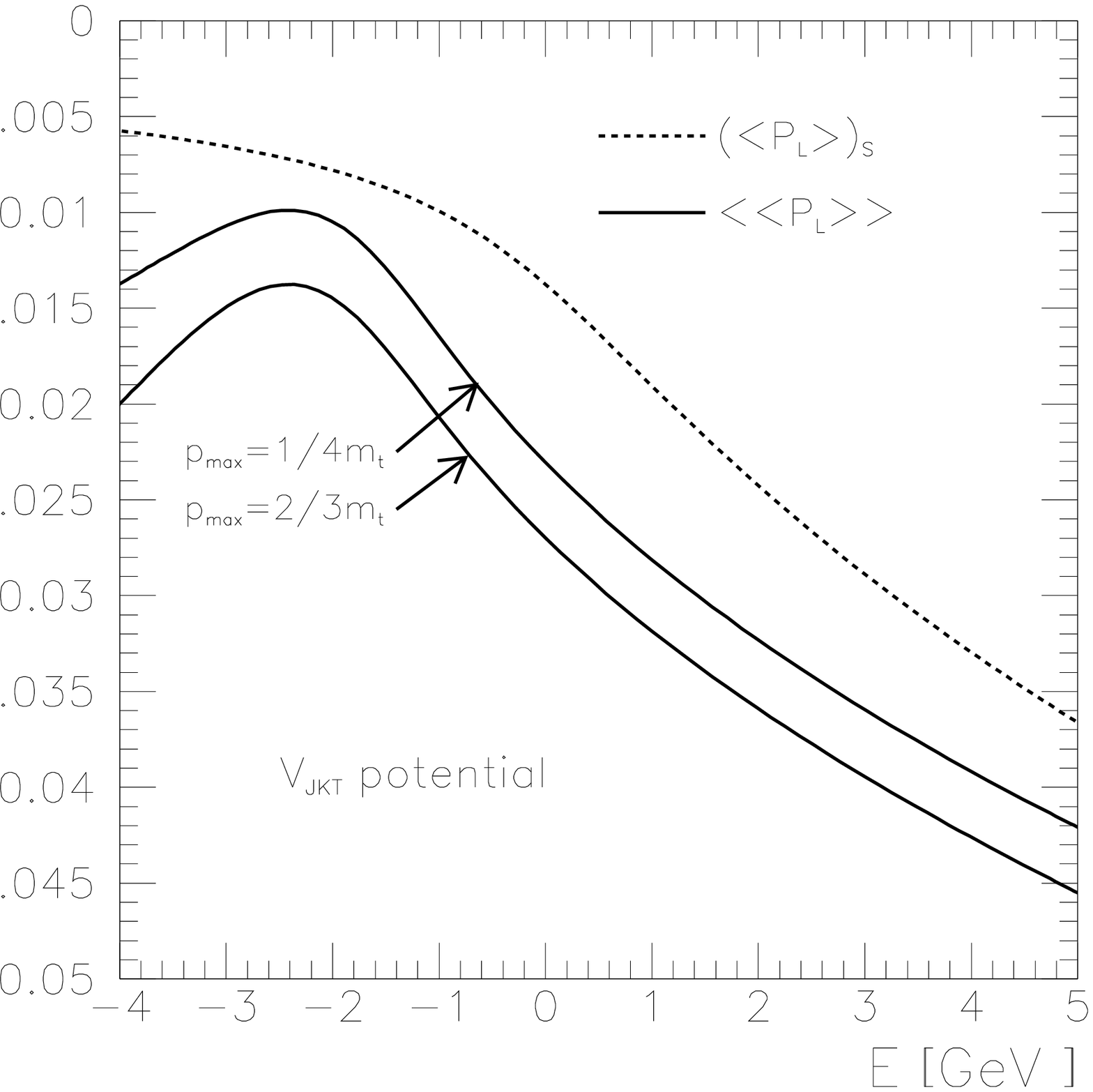,width=6.5cm,height=6.5cm}
\caption[fig3]{\footnotesize \it Angular and momentum average of
        the longitudinal top quark polarization 
	$ \langle\langle {\cal P}_L \rangle\rangle $ 
        for the pure Coulomb potential (left)
	and   $V_{JKT}$ (right) 
	shown in comparison with  
        $\left(\langle {\cal P}_L\rangle\right)_S$. }
\label{fig3}
\end{center}
\end{figure}
For a pure Coulomb potential and positive energies the formula (\ref{pola4}) can be 
reproduced, as shown in Fig.~\ref{fig2}a  and Fig.~\ref{fig3}a.
However for negative energies in the region of the 1S peak there are significant 
differences. 

A further improvement in the analysis can be obtained by including the phenomenological 
potential \cite{jkt}.
This is shown in Fig.~\ref{fig2}b and Fig.~\ref{fig3}b.
This time the normalization of the curves is significantly different even in 
the positive energy region.

We think that the polarization at 1S peak is sensitive to both, the top quark width 
$\Gamma_t$ and $\Delta E_{1S-2P}$ energy difference between 1S and 2P peak values. The 
formula (\ref{pola4}) is not sensitive to  $\Delta E_{1S-2P}$ and hence in our opinion 
cannot be considered a very good approximation.

\section{Rescattering corrections}

Due to the extremely short life time of the top quark it will never be possible
to carry out any experiment directly on it. 
The most suitable way of determining the observables discussed here is based on the analysis of
charged leptons from the semileptonic decay channel: 
$
e^+\,e^- \rightarrow  t\,\overline{t} \rightarrow b\, l \,\nu \,\overline{b}\, W^-. 
$ 
The average of the charged lepton distribution is relatively easy to settle theoretically 
as well as experimentally 
\begin{equation}
  \langle nl\rangle \equiv
       \Big(\frac{d^3\sigma}{dp\, d\Omega_p}\Big)^{-1}\int dE_l\, d\Omega_l
       \frac{d\sigma(e^+e^-\to bl\nu\bar bW^-)}{d p\, d\Omega_p\, d\Omega_l\, d E_l}
       (nl) \, ,
\end{equation}
where {\em l} is the four-momentum of the charged lepton and 
{\em n} is a chosen unit four-vector. 
In \cite{HJKP} it was calculated  as 

\begin{equation}
 \langle nl\rangle = \mbox{BR}(t\to bl\nu) \frac{1+2y+3y^2}{4(1+2y)}
    \Big[ (tn)+\frac{m_t}{3} \, (n{\cal {\cal P}}) \Big] .
\end{equation}
In the $t\bar t$ center-of-mass frame $ t^\mu = (m_t, {\bf p})$ is the top 
quark four-momentum and $ y = \frac{m_W^2}{m_t^2}$.   
The components of ${\cal P}$ turn out to be 
identical with those of the top quark polarization 
originating from the S-P wave interference 
in the \mbox{$e^+e^- \rightarrow t\overline{t}$} process.
Thus we consider  ${\cal P}$ to represent the polarization.

In \cite{HJKP} it was also shown that the largest known corrections
to $\langle nl \rangle$
arise from the color interaction between bottom and antitop quarks 
(or top and antibottom).
An intuitive picture is given in \cite{mpys} and 
\cite{mjys}. 
Corrections due to this effect are called 'rescattering corrections', or 
'final-state corrections', or 'non-factorizable corrections'. 
It is known that they almost disappear in the total cross sections 
but they modify the differential distributions (\cite{sphd}, \cite{HJKP}).    
We implement them in our approach in the same way as we did before
(see Appendix A).
\begin{figure}[h!]
\begin{center}
\epsfig{file=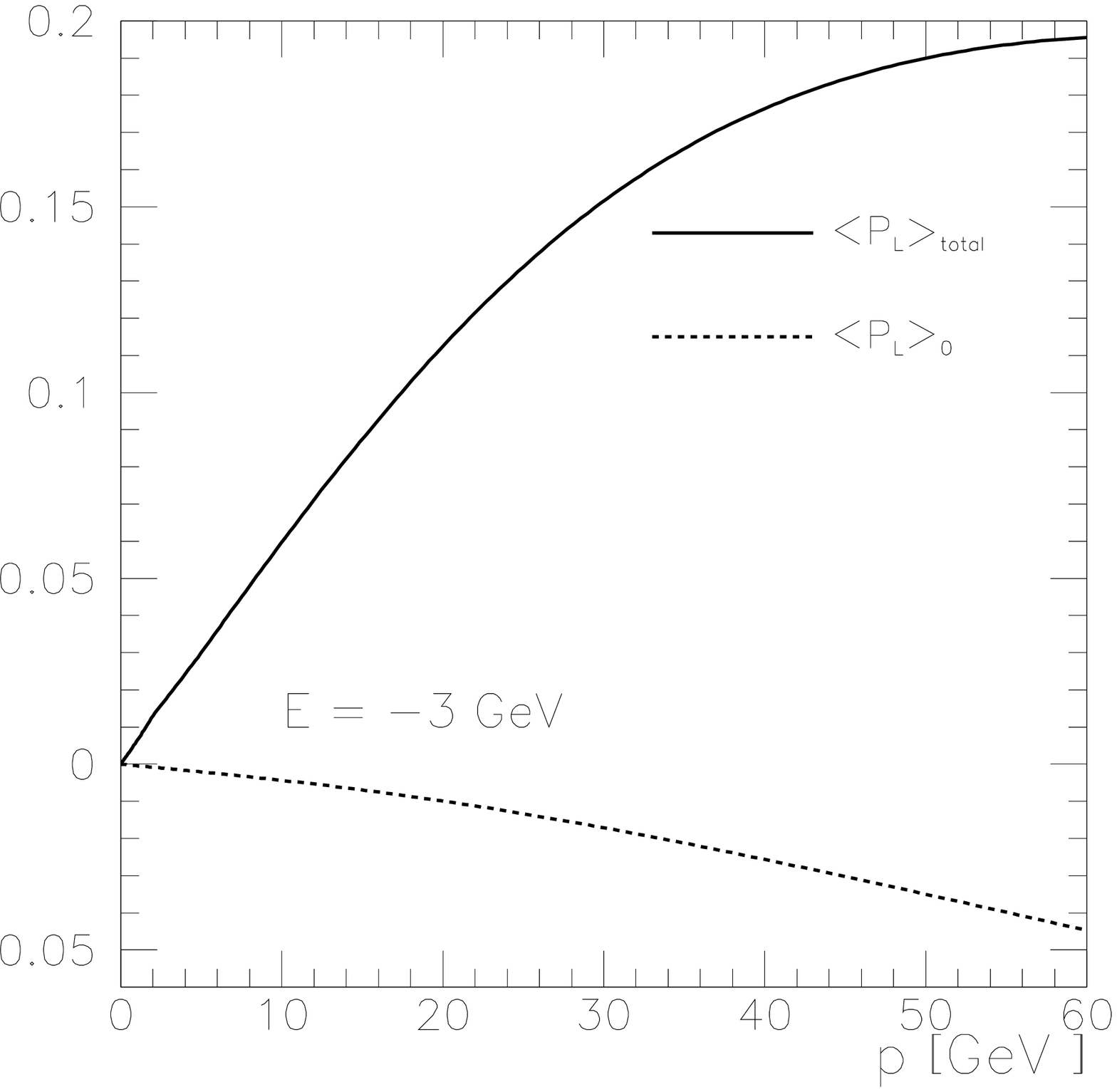,width=6.5cm,height=5.5cm}
\epsfig{file=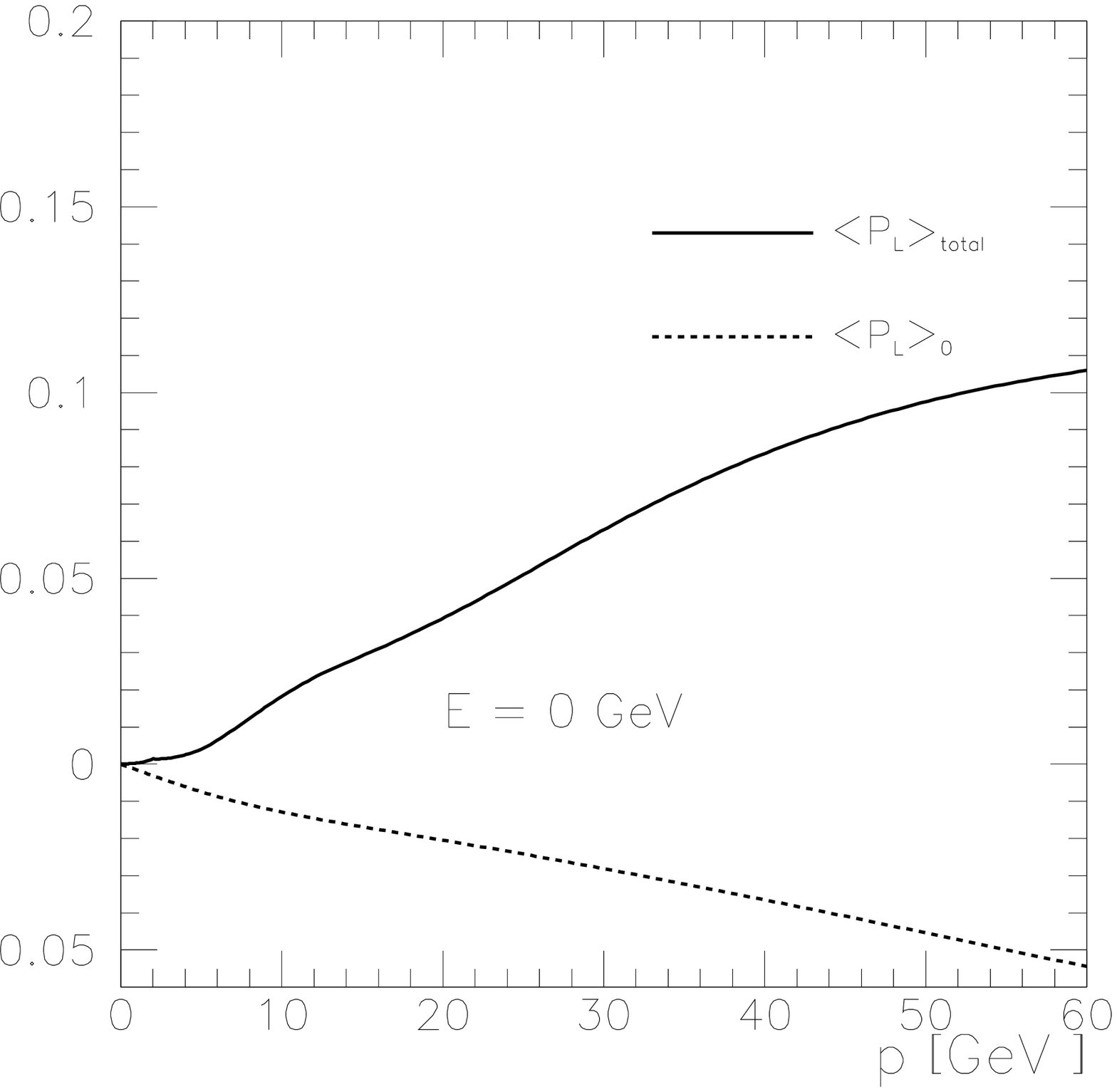,width=6.5cm,height=5.5cm}
\epsfig{file=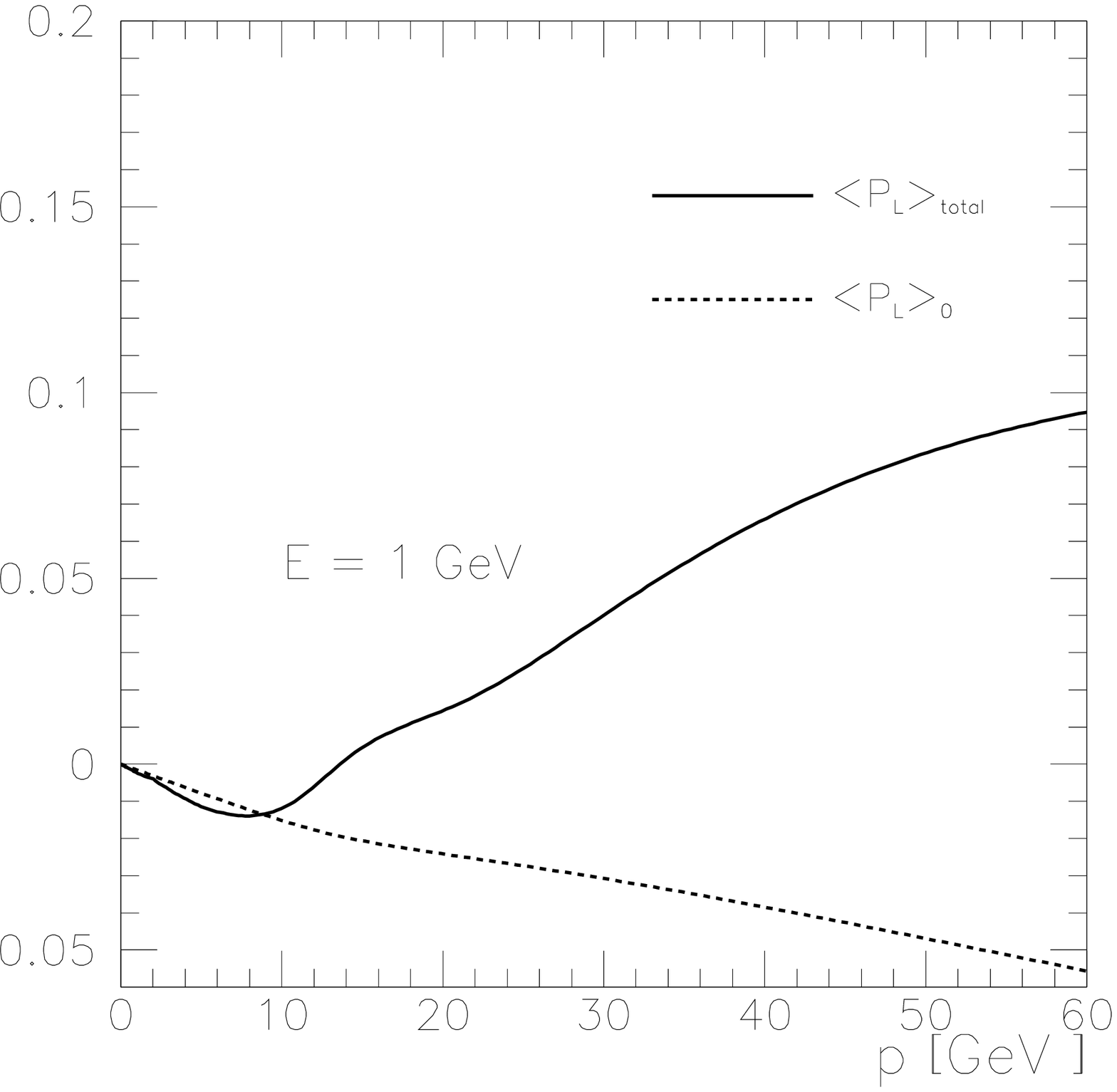,width=6.5cm,height=5.5cm}
\epsfig{file=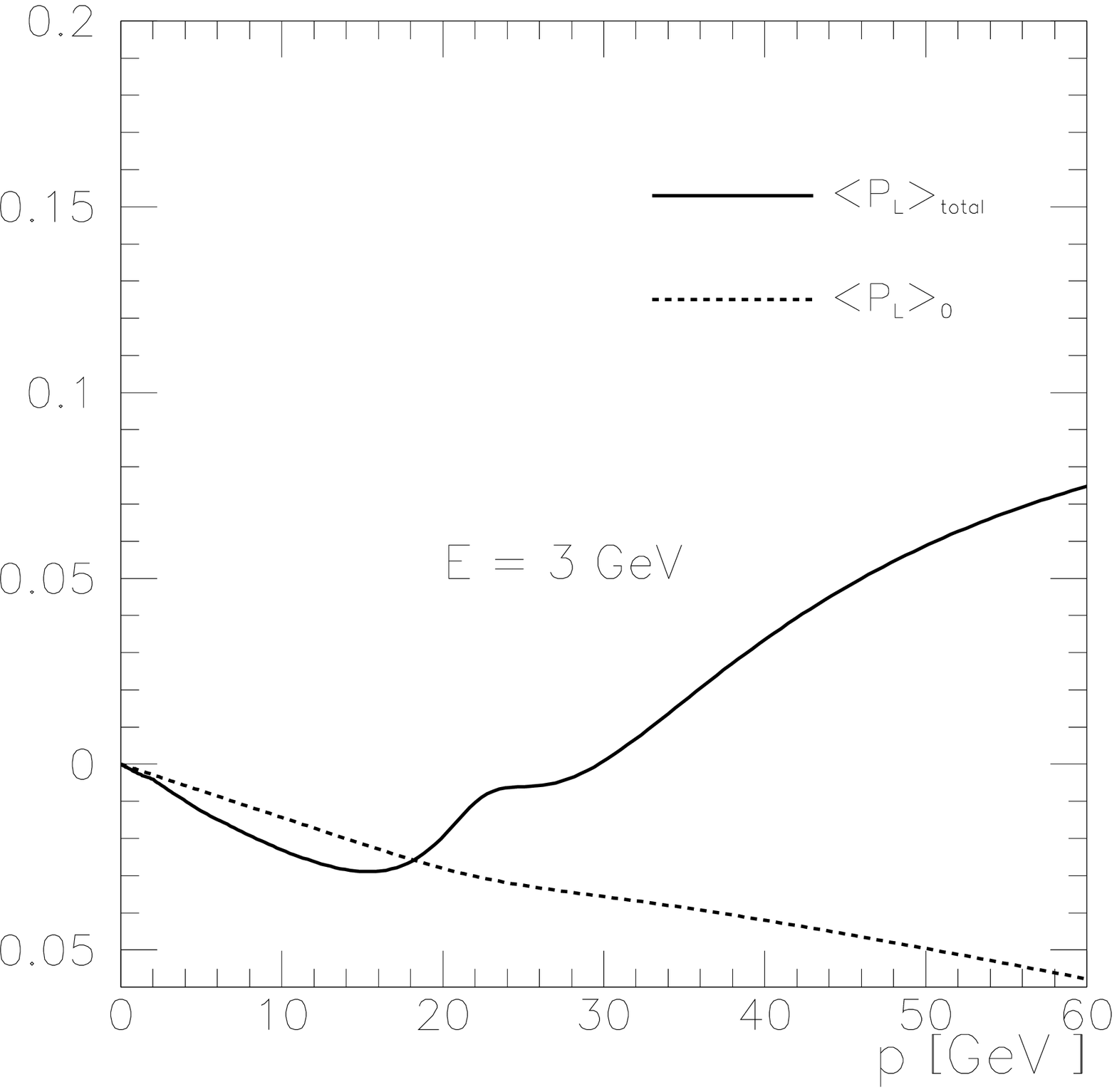,width=6.5cm,height=5.5cm} 
\caption{\footnotesize \it  Angular average of the 
        longitudinal top polarization with rescattering $ \langle {\cal P}_L\rangle$
        (solid line), compared with the S-P wave interference term
        (dotted line), for several different energies.}
\label{fig4}
\end{center}
\end{figure}
Then the  longitudinal top quark polarization, 
the polarization perpendicular to the top quark momentum 
and normal to the production plane averaged over the angles are respectively: 
\begin{eqnarray}
\langle{\cal P}_{L}\rangle &=& \frac{4}{3}\, C_{\bot} \, \varphi_R 
        +\frac{1}{3}\, k_{1}\, Re(\Psi_{2}) , \\
\langle{\cal P}_T\rangle &=& -\frac{\pi}{4}\, C_{||}^0 
        + \frac{\pi}{16}\, k_{2}\, C_{||}^0\, \Psi_{3}, \\
\langle{\cal P}_{N}\rangle &=& \frac{\pi}{4}\, C_{N}\, Im(\varphi) ,
\end{eqnarray}
where
\begin{eqnarray}
k_{1}&=&2\frac{3(1-3y^2)}{2(1+2y+3y^2)} 
     +\frac{2+3y-5y^2-12y^3}{(1+2y)(1+2y+3y^2)} ,\\
k_{2}&=&-2\frac{3(1-4y+3y^2)}{8(1+2y+3y^2)}
        -\frac{1-4y+3y^2}{4(1+2y+3y^2)} .
\end{eqnarray}
\begin{figure}[h!]
\begin{center}
\epsfig{file=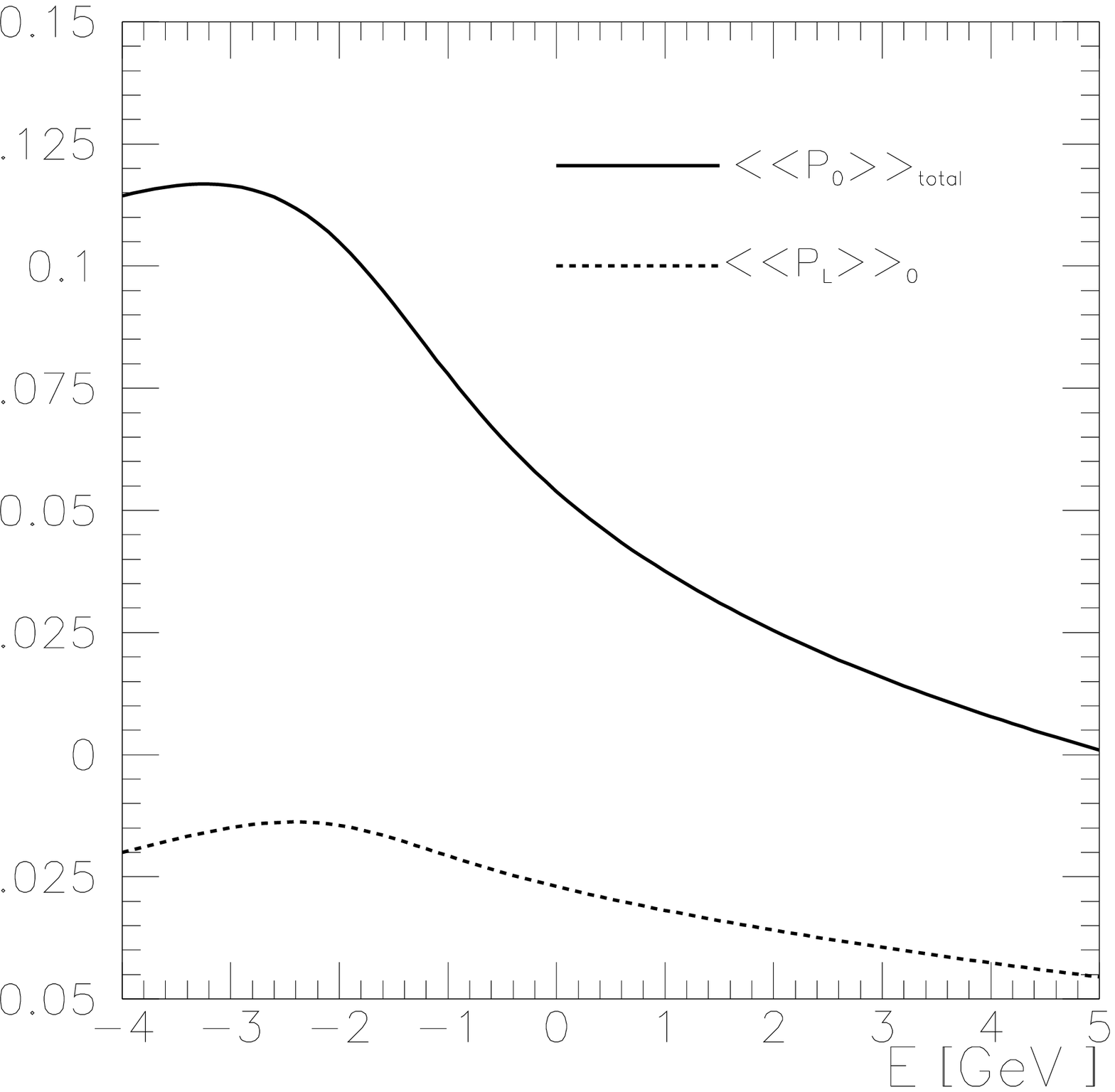,width=6cm,height=6cm} 
\caption{\footnotesize \it   Angular and momentum average of the 
        longitudinal top polarization with rescattering 
	$\langle\langle {\cal P}_L \rangle\rangle$ (solid line), 
        compared with the S-P wave interference term (dashed line).}
\label{fig5}
\end{center}
\end{figure}

The functions $ \Psi_{2} $ and $ \Psi_{3}$ (see ~\cite{HJKP}) are connected with 
the rescattering corrections in the $ t\,\overline{b}$ system  
and appear to be of the same order of magnitude. For $m_t = 175$~GeV
the coefficients $k_i$ are:    
$k_1 = 2.7 $ and $k_2 = -0.19$ 
so the corrections to the longitudinal polarization are proportional to 
$ 0.9 \, Re(\Psi_2)$ 
and to the perpendicular polarization:
$0.04\, \Psi_3$. 
Thus the perpendicular and the normal components 
of the polarization are almost unchanged by rescattering, but the helicity of 
the top quark is strongly affected.
\begin{figure}[h!]
\begin{center}
\epsfig{file=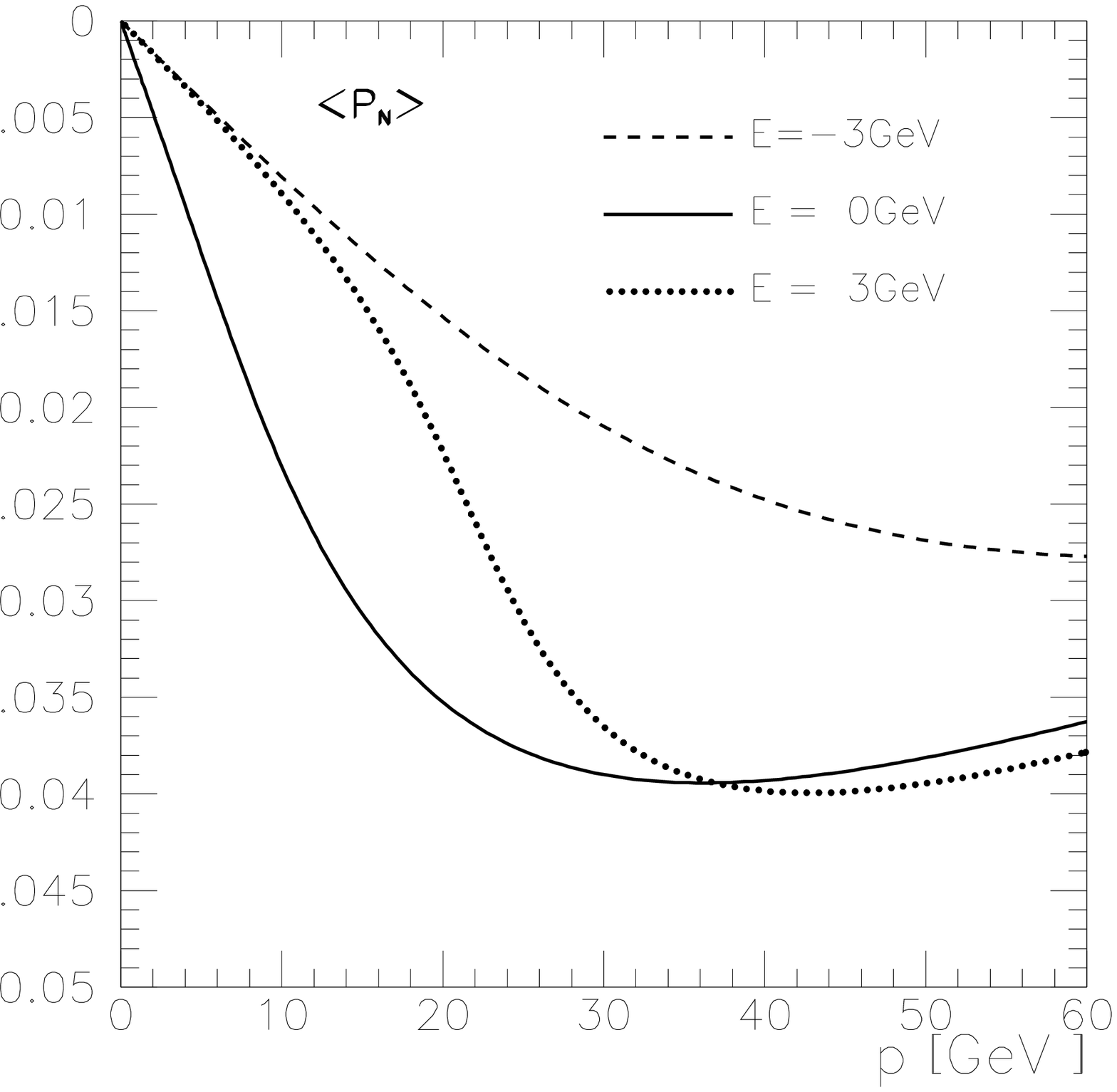,width=6.5cm,height=6.5cm} 
\epsfig{file=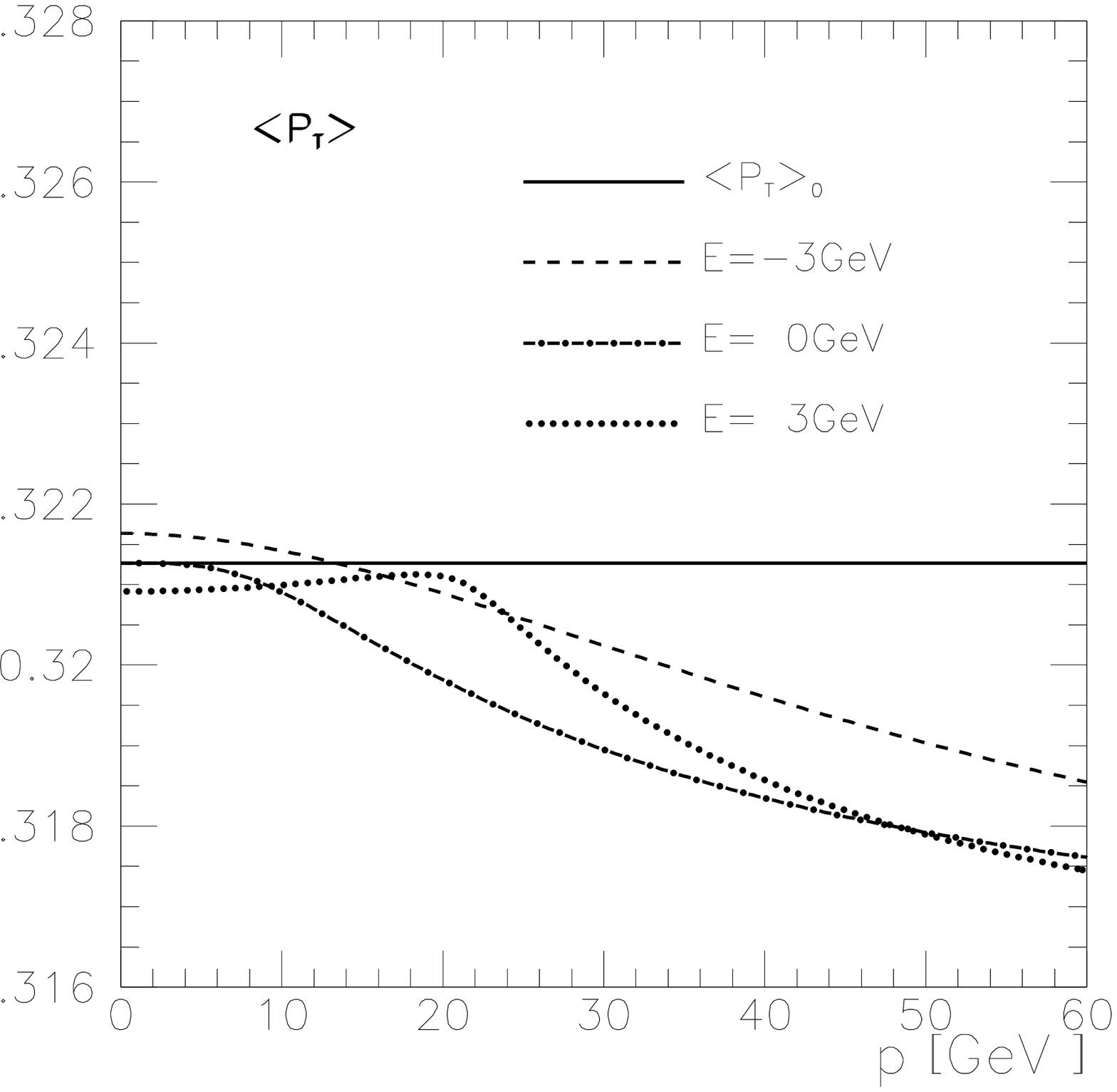,width=6.5cm,height=6.5cm} 
\caption{\footnotesize \it   Angular average of the 
        normal 
	$\langle {\cal P}_N \rangle$ (left) 
        and transverse 
	$\langle {\cal P}_T \rangle$ (right)
	top polarizations  
	plotted for several energies. 
        For $\langle {\cal P}_T \rangle$, the solid line 
	shows the result without rescattering 
        corrections (which does not depend on energy). The remaining
	represent the complete result.}
\label{fig6}
\end{center}
\end{figure}

We also consider the momentum average of the polarization
\begin{eqnarray}
\langle\langle{\cal P}_{L}\rangle\rangle &=& \frac{4}{3}\,C_\perp\,
                        \frac
                        {\int_{0}^{p_{max}}\,  
                        |p\, G(p,E)|^2 \, \varphi_R \, dp}
                        {\int_{0}^{p_{max}}\,         
                        |p\, G(p,E)|^2\,  dp} \\
                   &+& \frac{1}{3}\, k_{1} \,
                        \frac {\int_{0}^{p_{max}} \, 
                                |p\, G(p,E)|^2 \, Re(\Psi_2)\, dp}
                             {\int_{0}^{p_{max}}        
                                |p\, G(p,E)|^2\, dp} \nonumber ,\\
\langle\langle{\cal P}_T\rangle\rangle &=&-\frac{\pi}{4}\, C_{||}^0 + 
                    \frac{\pi}{16}\, k_{2}\, C_{||}^0\, 
                     \frac {\int_{0}^{p_{max}} 
                                |p\, G(p,E)|^2\, \Psi_3\, dp}
                             {\int_{0}^{p_{max}}\,        
                                |p\, G(p,E)|^2\,  dp} , \\
\langle\langle{\cal P}_{N}\rangle\rangle &=&\frac{\pi}{4}\, C_N\, 
                        \frac
                        {\int_{0}^{p_{max}}\,  
                        |p\, G(p,E)|^2\, Im(\varphi)\, dp}
                        {\int_{0}^{p_{max}}\,         
                        |p\, G(p,E)|^2\,  dp} .
\end{eqnarray}
\begin{figure}[h!]
\begin{center}
\epsfig{file=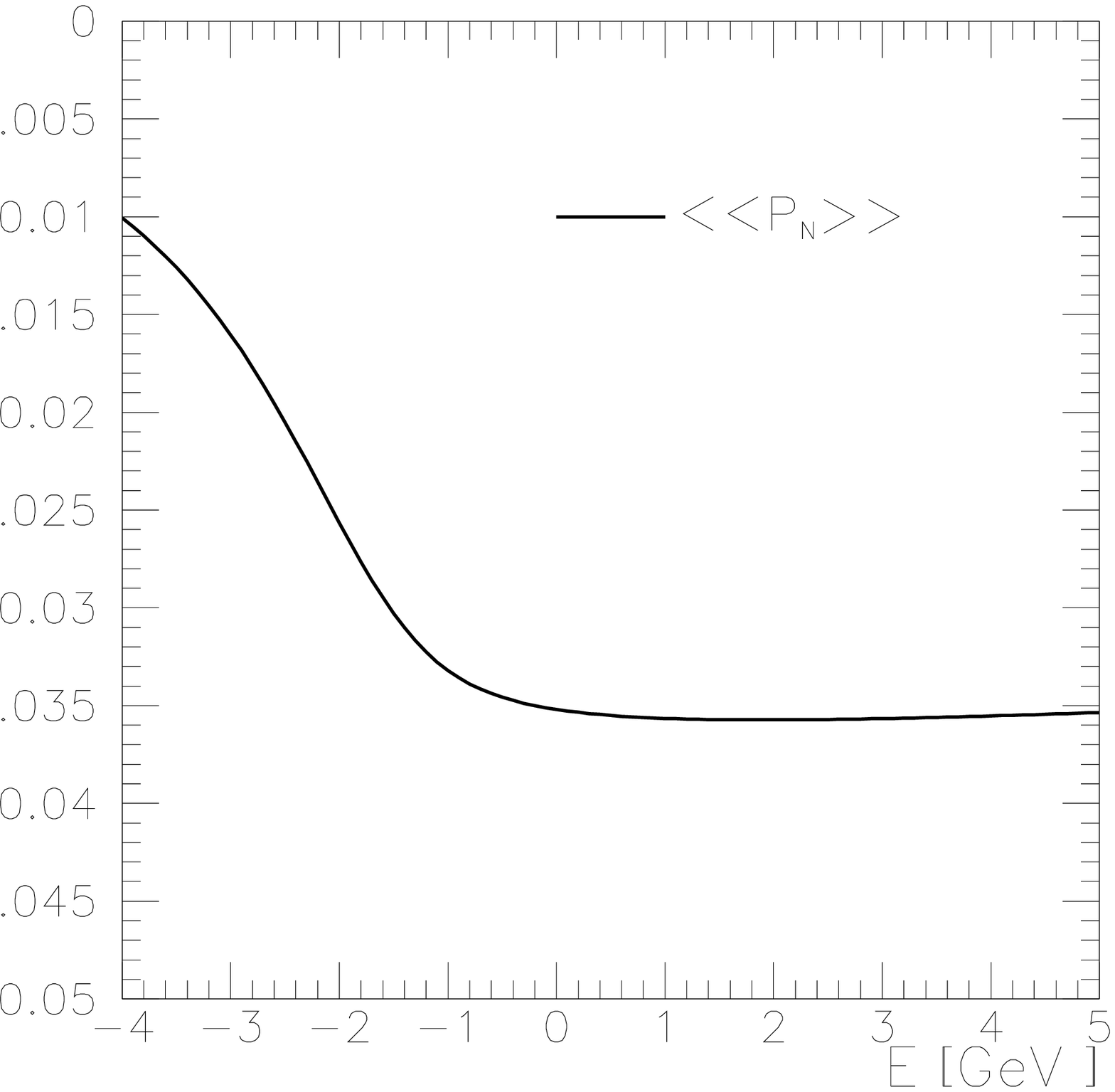,width=6.5cm,height=6.5cm} 
\epsfig{file=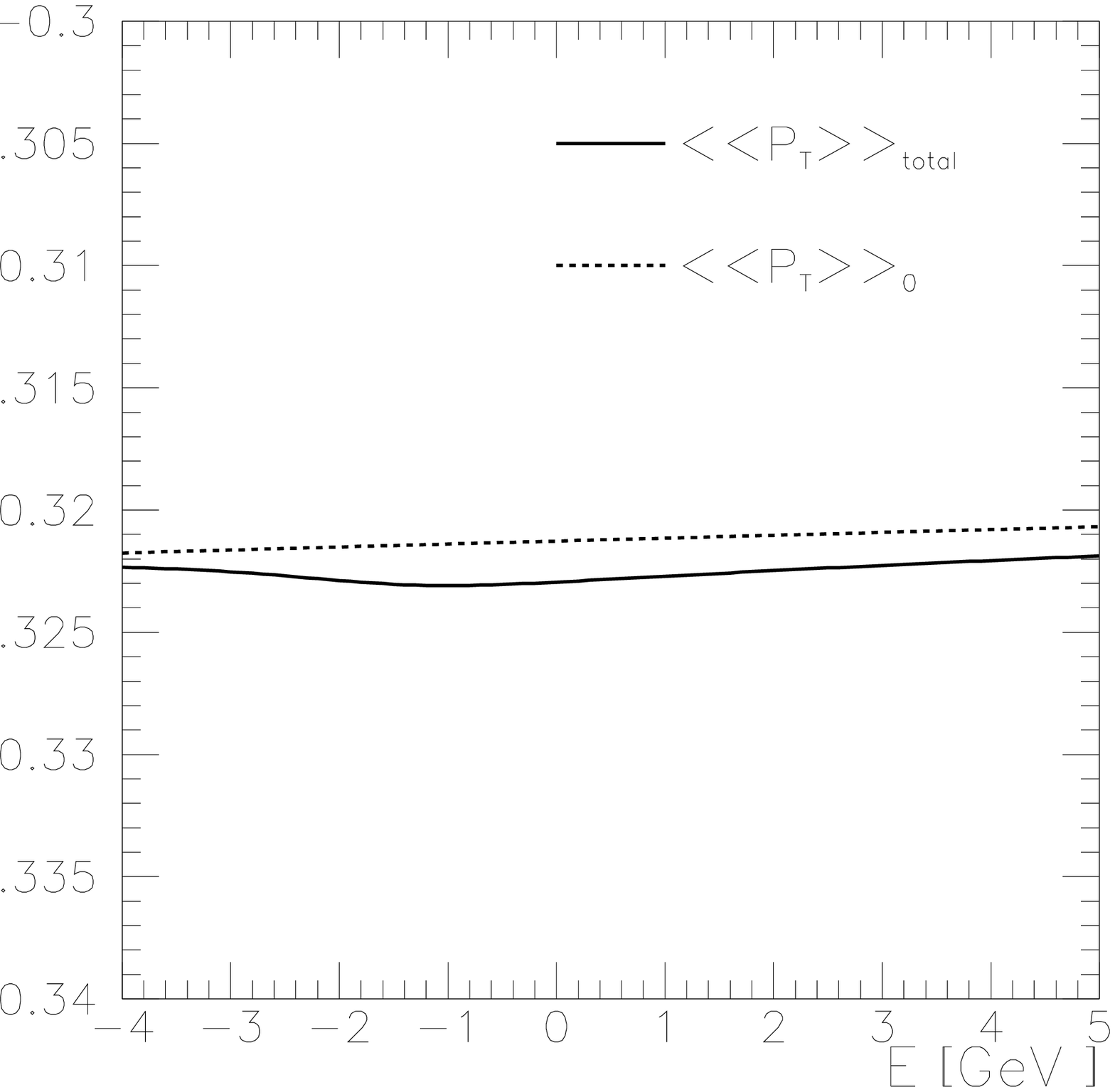,width=6.5cm,height=6.5cm} 
\caption{\footnotesize \it   
	Angular and momentum average of the normal
	$\langle\langle {\cal P}_N \rangle\rangle$ (left)
        and 
        transverse top polarization 
	$\langle\langle {\cal P}_T \rangle\rangle$ 
        (right). }
\label{fig7}
\end{center}
\end{figure}

In Fig.~\ref{fig4} we show the drastic change induced by the the final interactions on 
 ~\(\langle {\cal P}_L\rangle\). 
The quantity that we wish to compare with \cite{vol},~\(\langle\langle {\cal P}_L\rangle\rangle\)
is plotted in Fig.~\ref{fig5}. Again we see a substantial change in the normalization and the sign with respect to 
~\(\langle\langle {\cal P}_L\rangle\rangle\) from Fig.~\ref{fig3}.

For completeness we plot the normal and transverse components of the polarization in Fig.~\ref{fig6} and  
Fig.~\ref{fig7}.

\section{Conclusions}

We have shown that for negative energies relative to the threshold 
the approximation of \cite{vol} does not include the
main characteristics of the longitudinal polarization. Even if we neglect the final state
interactions the negative energy part is not reproduced. The phenomenological potential changes the normalization
slightly, leading to some differences in the positive energy part as well. 
But our non-relativistic approximation is too crude 
to seriously improve the predictions of ~\cite{vol} in this region. However a drastic difference appears
in a full analysis with the rescattering corrections. It should therefore be obvious that the top quark
parameters cannot be retrieved from a fit to Eq.~(\ref{pola4}) without the inclusion of the rescattering 
corrections.

Thus the average top quark longitudinal polarization may be a difficult quantity for phenomenological 
analysis aiming at determination of the top quark couplings. 
The size of the rescattering corrections should be viewed as an argument against the longitudinal
polarization as a source of top parameters. An interesting observable
weakly affected by the final state interactions is the forward-backward asymmetry.

\section*{Acknowledgments}

The authors wish to thank Thomas Teubner for his support at early stages of this work, 
helpful comments 
and careful reading of the manuscript. \\ 
This work is partly supported by the European Commission 5th Framework contract 
HPRN-CT-2000-00149 and by KBN grants 5P03B09320 and 2P03B05418.

\appendix
\section{Appendix}         
Here we would like to show how the results of \cite{HJKP} 
can be used to calculate the average top quark helicity. 

For the definition of the polarization Eq.~(\ref{pola}),
we need the differential cross section which is equal to    
\begin{equation}
 \frac{d\sigma}{dp \, d\Omega_{p}} d\Omega{_p} =
        \frac{\, d\sigma^{(0)}}{dp}\frac{1}{4\pi}
        (1+2\, A_{FB} \, \rm{cos}\vartheta \, ) \, d\rm{cos}\vartheta .  
\end{equation}
\( A_{FB} \) is the forward-backward asymmetry   
\begin{equation}
A_{FB}=C_{FB}\, \varphi_R .
\end{equation}
Then the averaged top polarization is 
\begin{equation}
\begin{Large}
\langle{\cal P}_i\rangle= \frac{1}{2} \int_{-1}^{1}{\cal P}_i \, 
			(1+2\, A_{FB}\, \rm{cos}\vartheta \, ) \, d\rm{cos}\vartheta .
\label{poli}
\end{Large}
\end{equation}
The unintegrated polarization distributions ${\cal P}_i$, which are 
the longitudinal ${\cal P}_L$, the transverse ${\cal P}_\perp$ 
and the normal polarization ${\cal P}_N$,
can be easily obtained from the previously calculated polarizations in the beam frame
\cite{HJKP} by a rotation of the basis by an  angle $\vartheta$
(the angle between top quark and electron momentum ) in the production plane.     
Then
\begin{eqnarray} \label{polhel}
 {\cal P}_{L} &=& {\cal P}_{\perp}^e \, \rm{sin}\vartheta
		+{\cal P}_{||}^e \, \rm{cos}\vartheta \nonumber ,  \\
 {\cal P}_{T} &=& {\cal P}_{\perp}^e \, \rm{cos}\vartheta
		-{\cal P}_{||}^e \, \rm{sin}\vartheta   ,\\
 {\cal P}_{N} &=& {\cal P}_{N}  \nonumber .
\end{eqnarray}
The index {\em e } was used to indicate that the quantities are given 
in the basis aligned with the electron momentum. 
We also use
\begin{eqnarray}
{\cal {\cal P}}_\|^e( p,E,\chi) &=& C_\|^0(\chi)
+ C_\|^1(\chi)\, \varphi_{\rm _R}(\rm p,E)\,\cos\vartheta\,
 \label{thr_long} , \\
{\cal P}_\bot^e( p,E,\chi) &=& C_\bot(\chi)\,
\varphi_{\rm _R}(\rm p,E)\,
\sin\vartheta\,
\label{thr_perp} ,\\
{\cal P}_{\rm N}^e( p,E,\chi) &=& C_{\rm N}(\chi)
\varphi_{\rm _I}(\rm p,E)
\sin\vartheta\,
        \label{thr_norm} .
\end{eqnarray}
The needed coefficients are ~\cite{hjkt,HJKP} \\
\parbox{75.ex}{
\begin{eqnarray*}
& &\hspace{5.ex}C_\|^0 (\chi) =
-{a_2 + \chi a_1 \over a_1 + \chi a_2} ,\hspace{6.9ex}
  C_\|^1 (\chi) = \left( 1-\chi^2 \right) {a_2 a_3 - a_1 a_4   \over
        \left(a_1 + \chi a_2 \right)^2} ,\\
& &\hspace{5.ex}C_\bot(\chi)  = -{1\over 2} \,
{a_4 + \chi a_3 \over a_1 + \chi a_2} ,
    \qquad C_{\rm N}(\chi) =-{1 \over 2}\, {a_3
    + \chi a_4 \over a_1 + \chi a_2}\, =\, - C_{\rm FB}(\chi) ,
\end{eqnarray*}}
\hfill
\parbox{5.ex}{
\begin{eqnarray} \label{coefs} \end{eqnarray} }
with
\begin{eqnarray}\label{aai}
a_1 &=& q_e^2 q_t^2 + (v_e^2 + a_e^2) v_t^2 d^2 +
        2 q_e q_t v_e v_t d \nonumber \\
a_2 &=& 2 v_e a_e v_t^2 d^2 + 2 q_e q_t a_e v_t d \nonumber \\
a_3 &=& 4 v_e a_e v_t a_t d^2 + 2 q_e q_t a_e a_t d \label{coupl}\\
a_4 &=& 2 (v_e^2 + a_e^2) v_t a_t d^2 + 2 q_e q_t v_e a_t d \nonumber\\
d &=& {1\over 16 \sin^2\vartheta_{\rm W}\cos^2\vartheta_{\rm W}}\,{s\over s - M_Z^2}.
    \nonumber
\end{eqnarray}
The following conventions for the fermion couplings are used: 
 $ v_f = 2\, I^3_f - 4 \, q_f \, \sin^2\vartheta_{\rm W} ,\,  a_f = 2\, I^3_f $.
The parameter $\chi$ can be interpreted as the effective longitudinal polarization of
the virtual intermediate photon or $Z$ boson:\\
$ \chi={{\cal P}_+-{\cal P}_-\over1-{\cal P}_+{\cal P}_-} $
where ${\cal P}_\pm$ denotes the longitudinal electron/positron
polarization.

In an analogous and quite straightforward way 
the rescattering corrections can 
be incorporated in our approach;
we only have to shift the polarizations ${\cal P}_i \rightarrow {\cal P}_i + \delta {\cal P}_i$ 
and the forward-backward asymmetry, as given in \cite{HJKP}. The results are 
shown in section 2. 

%
%
%
%

\def\yadfiz#1#2#3{{\it Yad.~Fiz.~}{\bf #1} (#2) #3}
\def\jetp#1#2#3{{\it JETP~Lett.~}{\bf #1} (#2) #3}
\def\app#1#2#3{{\it Acta~Phys.~Polonica~}{\bf B #1} (#2) #3}
\def\zpc#1#2#3{{\it Zeit.~Phys.~}{\bf C #1} (#2) #3}
\def\prd#1#2#3{{\it Phys.~Rev.~}{\bf D #1} (#2) #3}
\def\plb#1#2#3{{\it Phys.~Lett.~}{\bf B #1} (#2) #3}
\def\mpl#1#2#3{{\it Mod.~Phys.~Lett.~}{\bf A #1} (#2) #3}

\end{document}